\documentclass[aps,twocolumn,prl,superscriptaddress,reprint]{revtex4}
\usepackage{amsmath,graphicx,color,enumitem,hyperref}
\usepackage[FIGTOPCAP,raggedright,nooneline,bf,footnotesize]{subfigure}
\usepackage{overpic}
\usepackage[version=4]{mhchem}

\newcommand{\figref}[1]{Fig.~\ref{#1}}

\begin{document}
	
	\title{Dispersion measurement method with down conversion process}
	
	\author{Marta Misiaszek $^\dagger$}
	\author{Andrzej Gajewski $^\dagger$}
	\author{Piotr Kolenderski}
	\affiliation{Faculty of Physics, Astronomy and Informatics, Nicolaus
		Copernicus University, Grudziadzka 5, 87-100 Toru\'{n}, Poland}
	\email{kolenderski@fizyka.umk.pl}
	
	\begin{abstract}
		Proper characterization of nonlinear crystals is essential for
		designing single photon sources.  We show a technique for dispersion
		characterization of a nonlinear material by making use of phase matching
		in the process of parametric down conversion. Our method is demonstrated
		on an exemplary  periodically poled potassium titanyl phosphate
		\ce{KTiOPO4} crystal phase-matched for $396$ nm to $532$ nm and $1550$
		nm. We show a procedure to characterize the dispersion in the range of
		$390$ to $1800$ nm by means of only one spectrometer for the UV-visible
		range.
	\end{abstract}
	
	\maketitle

%\section{Introduction}
Single photon sources are essential for experimental implementations of various quantum information processing and communication protocols. One of the most popular types of such sources is based on the process of spontaneous parametric down-conversion (SPDC). The process takes place in a nonlinear medium where a photon of a pump beam decays into a pair of photons. Technically, information about dispersion in a crystal is the prerequisite for the design and fabrication of such a source. There are many crystals that allow for efficient pair generation.  Consequently, there is a plethora of possibilities for the manipulation of photon states \cite{Laudenbach2017, Kolenderski2009}. Those properties are defined by crystal and pump characteristics, of which dispersion is the most important.

Dispersion is described by a tensor of rank two composed of electrical permittivities. After diagonalization, which is related to a respective rotation of the coordinate system, its diagonal elements are interpreted as refractive indexes \cite{SalehTeich2007,Hecht2002,Boeuf2000}. The related axes are traditionally called principal axes.  Typically, crystals used for photon pair generation  are uniaxial or biaxial, which means that the dispersion relation is described by two or three (respectively) independent elements of refractive index tensor. It results in a direction and polarization dependent velocity. This, in turn, is the basis for phase matching in nonlinear processes such as: spontaneous parametric down conversion, second harmonic generation (SHG), sum frequency generation (SFG) etc. 

The dispersion relation in a medium can be derived from first principles assuming a simple harmonic oscillator model. The resulting formulas are traditionally called the Sellmeier equations \cite{Marcuse1980a}. The form of the original equations was modified over time in order to better conform to experimental observations. Typically, the measurement technique which allows to determine the respective coefficients of the equations is based on a technique resorting to the process of SHG or SFG \cite{SalehTeich2007}. It was used for characterization of nonlinear crystals i.~e.~: \ce{LiInS2} \cite{Kato2014a}, \ce{BaGa4Se7} \cite{Kato2017},  \ce{GaS_xSe_{1-x}} \cite{Kato2014,Takaoka1999}, \ce{LiGaSe2} \cite{Miyata2017} etc.

In this letter we show a technique which is based on the nonlinear process of SPDC. Our method can be applied to various types of crystals assuming a phase matching characterization method is available \cite{Akbari2013,Schmidt2015,Riedel2014}.  It allows to determine the coefficients of the Sellmeier equations even with limited  detection spectral range. We demonstrate our method experimentally using a periodically poled \ce{KTiOPO4}  (PPKTP) crystal. The dispersion and temperature dependence for this crystal were analyzed before in Refs.~\cite{Kato2002,Manjooran2012,Stoumbou2013,Lee2012,Zhao2010a}. Our crystal is phase matched for $396$ nm to $532$ nm + $1550$ nm and can be tuned with temperature and pump wavelength. Using only one spectrometer for the range of $340$-$680$ nm, we show a procedure to determine the dispersion in the infrared range.
% In our experiment all waves are  approximately collinear and are polarized along slow polarization. This means that waves are practically z-polarized, hence only dispersion of one element of refractive index tensor can be measured, namely  $n_{z}$.

%\section{Quasi phase matching}
%{\color{red} Ta sekcja mi sie juz podoba, ale nie jestem pewna jak ma sie jej wartosc merytoryczna.}
%\comp{Przeredagowac.Logika sekcji: nieliniowos efektywna najwieksa -> jak spelnic dopasowanie fazowe -> sturkutra periodyczna->dyspersja> duza zalenos od syspersji-> wiec trzeba dobrze okreslic wspolczynniki. }

The pair production rate in the SPDC process depends on the effective nonlinearity of a medium and phase matching conditions \cite{Louisell1961, Boeuf2000}. Here, we use a reference frame of the principal axis of the crystal where the collinear propagation is along the x-axis. In general, the direction which results from the collinear phase matching condition  does not coincide with the direction for which the effective nonlinearity is the largest. In order to achieve efficient photon generation, crystals are periodically poled, which means that crystal's consecutive domains have the same absolute value of effective nonlinearity, but of opposite sign \cite{Fejer1992}. These leads to a quasi-phase matching (QPM) equation, which takes the form:
\begin{equation}
	\Delta \vec{k} =\vec{k_P}-\vec{k}_{\text{VIS}} -\vec{k}_{\text{IR}} - \frac{2\pi} {\Lambda} \hat{x}=\vec{0},
	\label{eq:QPM}
\end{equation}
where $\Lambda$ is a poling period of a crystal and $\vec{k_{P}},\vec{k_{VIS}},\vec{k_{IR}}$ are wavevectors of the pump, visible (VIS) and infrared (IR) photons, respectively. The momentum of a photon in a given mode is proportional to the effective refractive index, which involves, in general, all three indexes of refraction. The dispersion formula for each of the elements is given by the Sellmeier equations \cite{Marcuse1982}:
\begin{equation}
n_{j}^2(\lambda)=a_{j0}+\frac{a_{j1}}{\lambda^2-a_{j2}}+\frac{a_{j3}}{\lambda^2-a_{j4}}, \quad j = x,y,z,
\label{eq:Sellmeier}
\end{equation}
where  $a_{ji}$ are the Sellmeier coefficients. This form of the Sellmeier equation can be found in i.e.~Ref.~\cite{Kato2002}. In our case the pump photon propagates along the x-axis and its polarization is along the z-axis. The VIS and IR photons are slow polarized and their wavevectors create very small opening angle with wavevector of the pump photon.
Consequently, the dispersion in our example is determined by fifteen coefficients.

\begin{figure}[t]
	\centering
	\begin{overpic}[width=\columnwidth, clip]{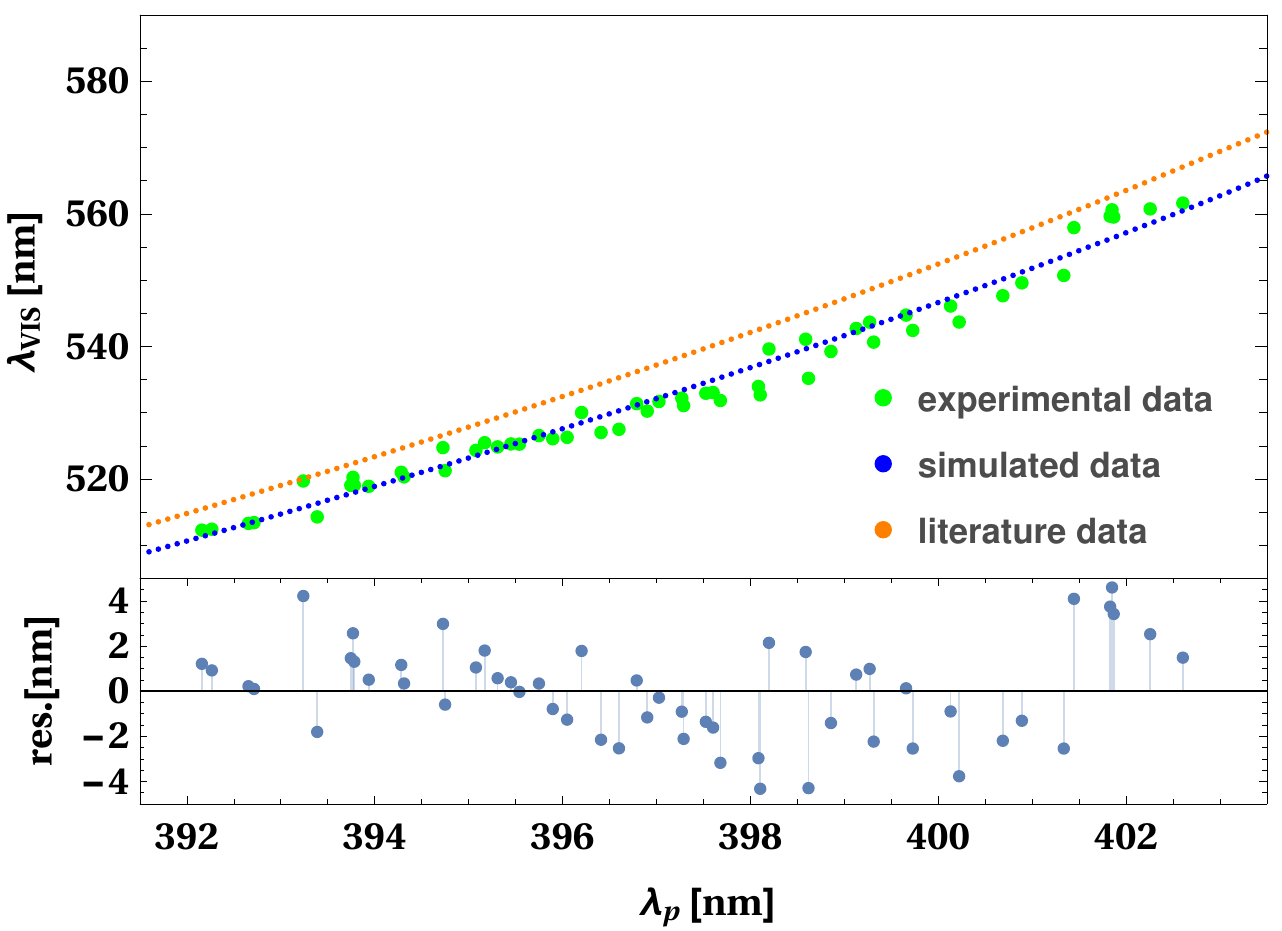}
		\put(9.5,71.5){
			\subfigure{\begin{tabular}{c}
					\includegraphics[width=0.38\columnwidth]{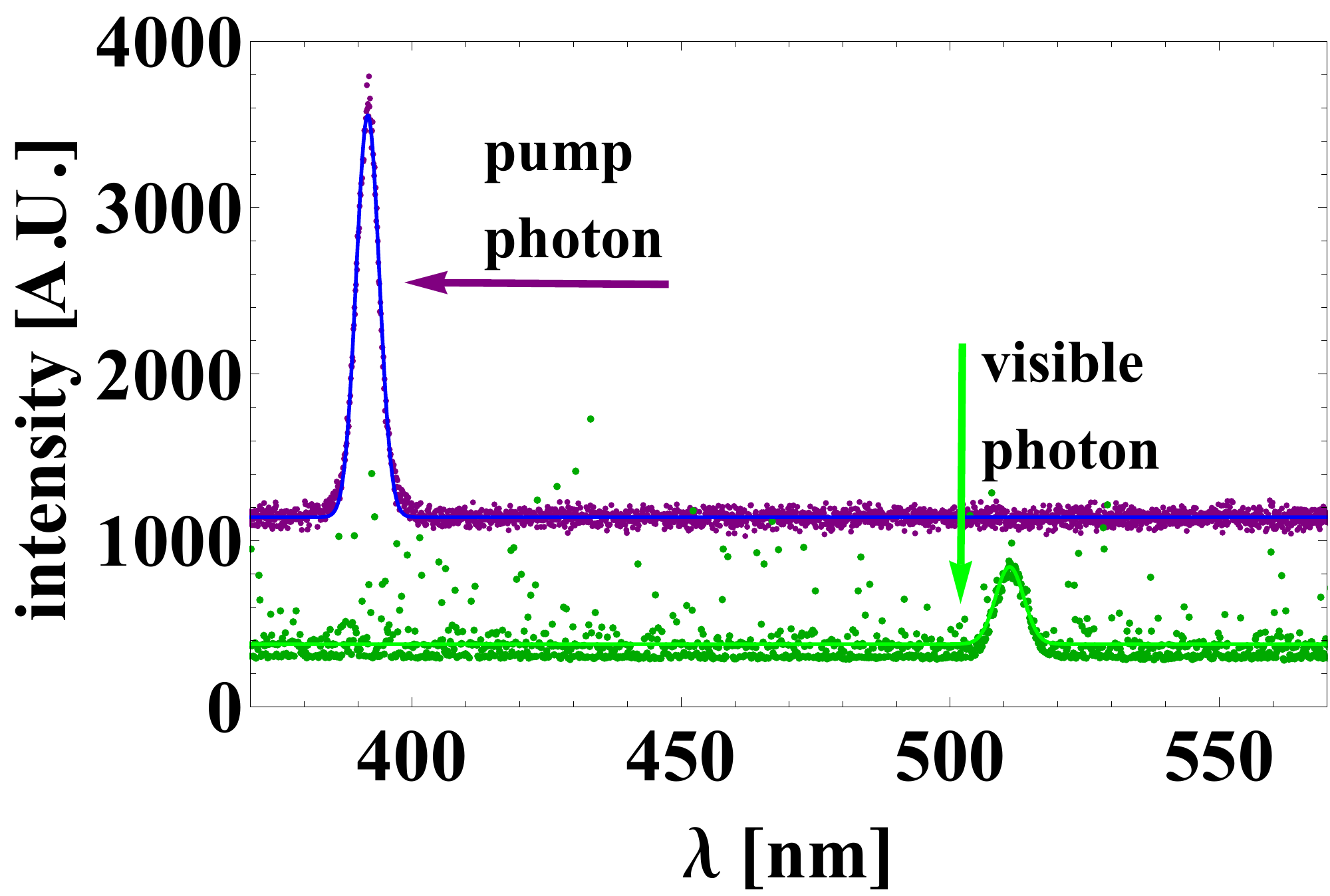}
				\end{tabular}}}
			\end{overpic}
			\caption{Experimental results of a visible photon wavelength dependence on the pump wavelength. Inset: an exemplary measurement of a visible photon and pump photon spectra. }
			\label{fig:pump}
\end{figure}

In order to calculate a central wavelength of out-coming photons, one needs to solve quasi-phase matching problem given in \eqref{eq:QPM} for photons inside a crystal. 
%We start from description of normal biaxial nonlinear crystal presented in Ref. ~\cite{Boeuf2000}. We are trying to find solution for periodically poled crystal, where one of the parameters is a direction outside the crystal (direction of measurement). Periodic polling modifies a phase matching equation by addition of pseudo vector term $\frac{2 \pi}{\Lambda}$ in \eqref{eq:QPM} and also by changing relation between wavevectors of visible and infrared photons since no longer those two have to add up to wavevector of pump photon. With that in mind, 
It can split into two separate equations -- one for wavevector components along the x-axis and one for perpendicular ones \cite{Boeuf2000}:
\begin{equation}
\begin{aligned}
n_{\text{VIS}} \frac{\omega_{\text{VIS}}}{\omega_{\text{IR}}} \sin(\theta_{\text{VIS}}) &= n_{\text{IR}}\sin(\theta_{\text{IR}}), \\
n_{\text{VIS}}\frac{\omega_{\text{VIS}}}{\omega_{\text{IR}}} \cos(\theta_{\text{VIS}}) &= n_p \frac{\omega_p}{\omega_{\text{IR}}} - n_{\text{IR}}\cos(\theta_{\text{IR}}) -\frac{2\pi} {\Lambda},
\end{aligned}
\label{eq:wav}
\end{equation}
where $\theta_{\text{IR}}$ and $\theta_{\text{VIS}}$ are angles created by wavevectors of IR and VIS photon with x-axis. These angles are defined inside the crystal and they relate to angles outside the medium by Snell's law. The analytical formula for IR photon angle $\theta_{\text{IR}}$ can be derived from \eqref{eq:wav}. In general, the VIS photon wavelength,
\begin{equation}
\lambda_{\text{VIS}}=\lambda_{\text{VIS}}(\omega_P,\Theta_P,\phi_P,\Theta_{\text{VIS}},\phi_{VIS},T,\Lambda_0,\vec{S}),
\label{eq:para}
\end{equation}
is a  function of the pump photon frequency $\omega_{P}$, angles of incidence of the pump photon wavevector with the surface of the crystal, $\Theta_P, \phi_P$, the position of the detector, $\Theta_{\text{VIS}}, \phi_{VIS}$, the temperature, $T$, the vector of the fifteen Sellmeier coefficients $\vec{S}$, and the length of periodic poling $\Lambda_0$.  This function can be only solved numerically. In our model the angular frequency of the pump photon $\omega_{P}$ is an argument for the algorithm and the wavelength of VIS photon is a returned value. Our method works in the following way.
The wavevectors, polarizations and wavelengths of pump and VIS photons are known from the experiment and Sellmeier coefficients can be easily found in the literature \cite{Kato2002,Manjooran2012,Stoumbou2013,Lee2012,Zhao2010a}. The  IR photon wavelength is determined by the pump and VIS photons wavelengths, because the three of them obey energy conservation relation. In order to numerically solve equations we used \textit{Wolfram Mathematica 11} software and its built-in function, \textit{FindRoot}. Method of finding a solution was set to Newton's method. All the arguments of the $\lambda_{VIS}$ function \eqref{eq:para} are assumed to be constant, with the exception of the angular frequency of the pump, $\omega_P$. In \figref{fig:pump} the solid red line shows the numerical solution of \eqref{eq:para} for our experimental setup settings using the Sellmeier coefficients from Ref.~\cite{Kato2002}. The experimental results are marked with green dots.

%\section{Measurement techniques}
\begin{figure}[h]
	\centering
	\includegraphics[width=\columnwidth]{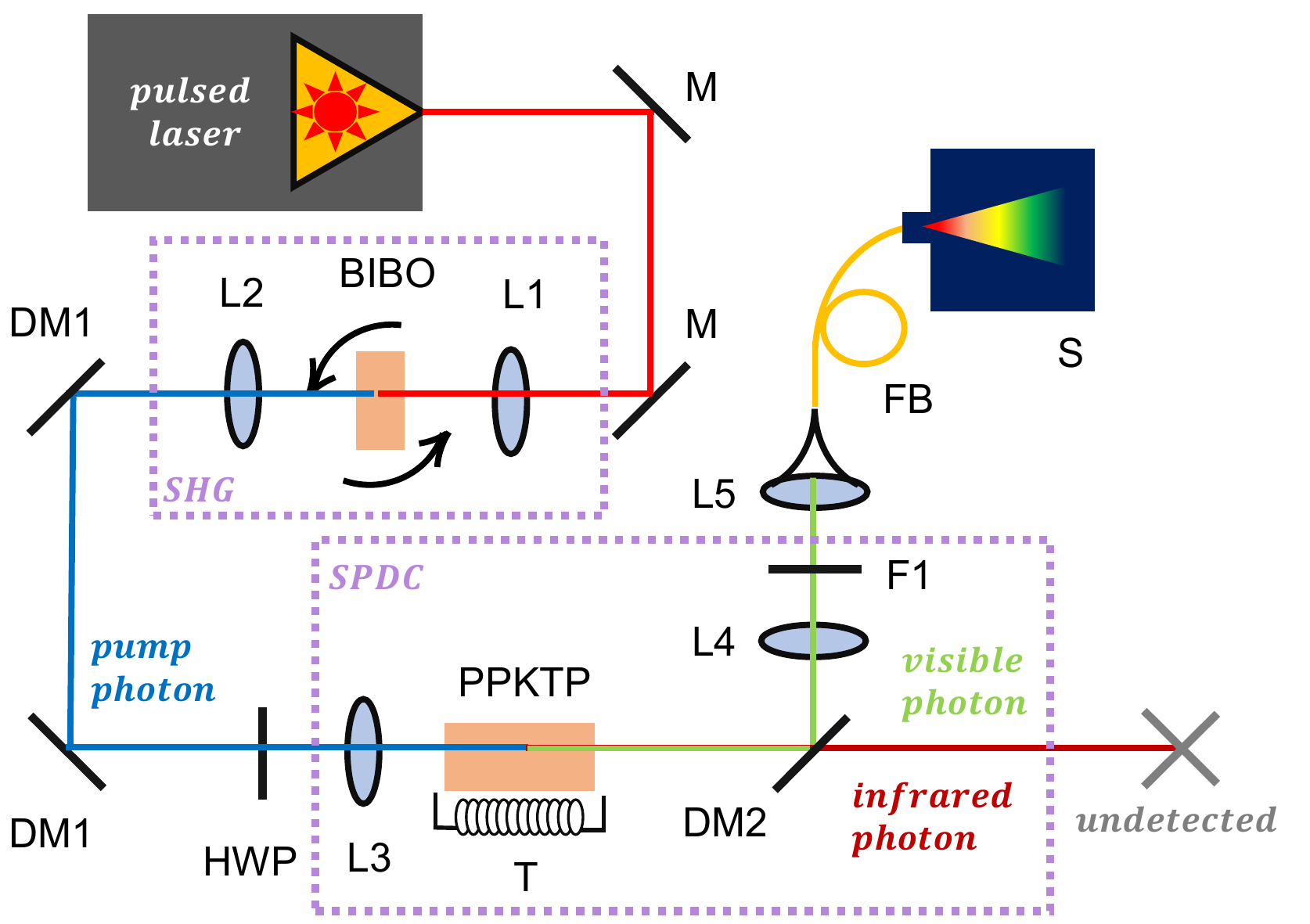}
	\caption{Experimental setup consists of: the pumping pulsed Ti:Sapphire laser, M -- mirror, L1, L2 --  lens (focal lengh f = 7.5 cm), BIBO -- bismuth triborate crystal, DM1 -- dichroic mirror (Semrock T425 LPXR), HWP -- half-wave plate, L3, L4 -- plano-convex lens (f = 10 cm, 12 cm), PPKTP --  periodically poled potassium titanyl phosphate crystal, T --  temperature controller, DM2 -- dichroic mirror (Semrock 76-875 LP), F1 -- set of filters (Chroma ET500, Z532-rdc ), L5 - aspheric lens (f = 1.51 cm), FB -- fiber (Thorlabs SMF460B), S -- spectrometer (Ocean Optics USB2000+).}
	\label{fig:setup}
\end{figure}

The experimental setup, depicted in \figref{fig:setup}, consists of two parts -- the second harmonic generation setup (SHG) and the SPDC source. A tunable femtosecond laser beam is frequency doubled in bismuth triborate (\ce{BiB3O6}) nonlinear crystal. A pair of dichroic mirrors DM1 separates the SHG and the laser beam. The frequency doubled beam is used to pump the PPKTP crystal, with a $4.01 \mu m$ poling period placed on a custom made mount T, which allows to control its temperature and position. The crystal temperature is kept constant at $304.8$ K.
The VIS photon of generated pair is in the spectral range of $500$ to $570$ nm and IR photon $1300$ to $1900$ nm. Next, the beam is split by dichroic mirror DM2, which transmits IR photons while reflecting the VIS and pump photons. Further, a set of filters, F1, reflect the pump photons and transmits the VIS photons. The VIS photons are coupled into a fiber, whereas the IR photons remain undetected.

The measurement procedure is as follows: the femtosecond  laser beam is set between $784$ nm and $806$ nm, which results in a pump photon wavelength in the range $392$ nm to $403$ nm. For each laser wavelength setting, the spectra of the VIS and the pump photons are measured using a spectrometer. The inset in \figref{fig:pump} presents an example spectra for a pump photon wavelength setting. Next, by fitting gaussian functions, the central wavelengths of both pump and VIS photons are obtained. The results are depicted in \figref{fig:pump} using green dots.

%\section{Results/Discussion}

The discrepancy between the experiment and analytical predictions is easy to see. This result was the motivation for further numerical analysis, namely: the computation of Sellmeier coefficients. The goal was to obtain coefficients which would describe more precisely the dispersion in PPKTP crystal in the spectral range of the pump, VIS and IR photons, which is approximately from $390$ nm to $1800$ nm. We don't want to significantly change the value of dispersion for longer wavelengths. In order to do that we fit the Sellmeier coefficients in  the model given in \eqref{eq:QPM} to our experimental data. We reduce the computational effort of numerical optimization by making a few observations. In our case, the dispersion depends predominantly on one element of a tensor, namely $n_z$. It is because of type $0$ configuration, where all the photons have the same polarization and which is close to the collinear propagation.  The values of the respective coefficients can be found in Tab.~\ref{tab:para} and Ref.~\cite{Kato2002}. The Sellmeier equations consist of a constant term and two resonant ones. These resonances correspond to $\sqrt{a_{z3}}=218$ nm and $\sqrt{a_{z5}}=9280$ nm, so only the first one is significant for our spectral range. Therefore, in our calculations, we fit only first three coefficients $a_{z0},a_{z1},a_{z2}$. The literature values \cite{Kato2002} are the starting point for our optimization procedure.
% Slight changes to these parameters don't influence dispersion for longer wavelengths.

% In the end  function \eqref{eq:para} can be rewritten as

%\begin{equation}
%\lambda_{\text{VIS}}=\lambda_{\text{VIS}}(\omega_P^N;a_{z0},a_{z1},a_{z2};\vec{c'}),
%\label{eq:para2}
%\end{equation}

%where measured pump photon angular frequency $\omega_P^N$ is an argument of function, Sellmeier coefficients of interest $a_{z0},a_{z1},a_{z2}$ are parameters of function and $\vec{S'}$ is a vector of constants.

In order to qualitatively compare Sellmeier equations with the literature (L) and our computed (C)  coefficients, we calculate the residual sum of squares (RSS) in the following way: 

\begin{equation}
\begin{split}
RSS^{j} = \sum_{i=1}^{N=55} (\lambda_{VIS}^i - \lambda_{\text{VIS}}(\omega_P^i; a_{z1}^{j},a_{z2}^{j},a_{z3}^{j};\vec{C}))^2, \\ j=L,C,
\end{split}
\label{eq:RSS}
\end{equation}

where $\lambda_{\text{VIS}}^i$ are the measured central wavelengths, which are depicted in \figref{fig:pump}, and $\vec{C}$ is a vector composed of all the model parameters with Sellmeier coefficients included, which we keep fixed: $\Theta_P= \frac{\pi}{2}$, $\Phi_P= 0$,  $\Theta_{VIS}= 0.01237$, $\Phi_{VIS}= 0$,$T = 304.86 K$, $\Lambda_0 = 4.01 \mu m$ ( in $298 K$).

In the first step we tested the validity of our model by comparing its outcomes with prediction of SNLO ~\cite{SNLO} software for collinear SPDC. It uses the Sellmeier coefficients from Ref.~\cite{Kato2002}. We got a perfect agreement. In the next step we used our model to generate a test data with random Gaussian noise up to $5 \% $. Then we used another built in function, NonLinearModelFit (NLF), with conjugate gradient method to estimate Sellmeier coefficients for the test data.  The NLMF numerically looks for values of parameters that minimize $RSS$, which quantify the quality of each individual fit. For $55$ data points, NLMF  was able to retrieve parameters with very high accuracy. These computations were also performed multiple times. At this point we established that our method works and can be trusted to compute Sellmeier coefficients which would best fit our data.

For the experimental data set of $55$ measurements, the NLMF starts from the initial Sellmeier coefficients, which we take from the Ref.~\cite{Kato2002}. Next, it computes new parameters for the next step and evaluates new $RSS$ value. The procedure stops when the specified accuracy is reached. We repeated that step $1000$ times and achieved the same result every time. That proves the stability of the  NLMF algorithm. We proceed with computation of the Sellmeier coefficients for the experimental data. We repeated that step $1000$ times to be certain of determinism of our method. The variances of coefficients  were below $2.1\times 10^{-3}\ \%$.

\begin{figure}[h]
	\centering
	\includegraphics[width=\columnwidth] {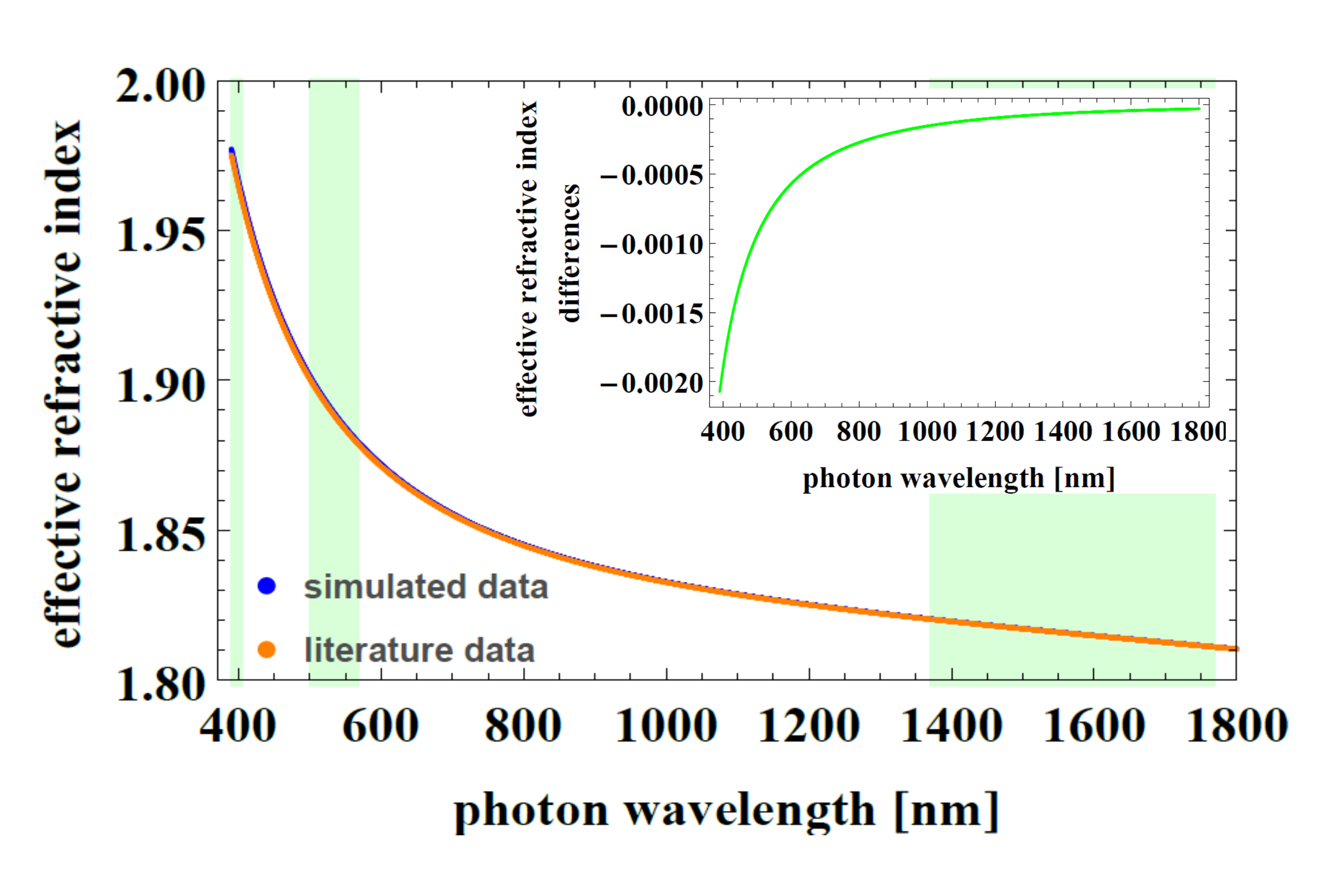}
	\caption{Effective refractive index. The shaded areas depict the wavelength range accessible by our method in our experiment. Inset: differences between literature and simulated data.}
	\label{fig:neff}
\end{figure}

The experimental, literature and simulated results are shown in \figref{fig:pump}. The comparison of calculated effective refractive index and literature one is presented in \figref{fig:neff}. Despite the differences being tiny, as seen in the inset of \figref{fig:neff}, change in photon wavelengths is significant. The residual sum of squares of fit for literature effective refractive index is equal to $RSS^L = 1910$ nm$^2$, whereas for our computed one it is $RSS^C = 257$ nm$^2$. Those values correspond to an average error of approximately $5.9$ nm and $2.2$ nm, respectively. We used that average error as a measurement error for each data point. As a result NLMF returned parameters with estimated uncertainties.  The values and uncertainties of the computed parameters are shown in Tab.~\ref{tab:para}.

%\begin{table}
%	\centering
%	\tabcolsep=0.11cm
%	\begin{tabular}{|c|c|c|} \hline 
%		Sellmeier  & literature   & computed   \\
%		 coefficient &  value  \cite{Kato2002} &  value   \\
%		\hline
%		$a_{z0}$ & $4.59423$ & $4.59423$ \\
%		$a_{z1}[\mu m^2]$ & $0.06206$ & $0.06274$\\
%		$a_{z2}[\mu m^2]$ & $0.04763$ & $0.04811$ \\ \hline
%		$RSS [nm^2]$ & $1900  $ & $250 $ \\ \hline
%	\end{tabular}
%	\caption{Comparison of the literature and calculated Sellemeier coefficients as in \eqref{eq:Sellmeier}}.
%	\label{tab:para}
%\end{table}

\begin{table}
	\centering
	\tabcolsep=0.11cm
	\begin{tabular}{|c|c|c|c|c|c|c|} \hline 
		  & $a_{z0}$   & $a_{z1}$ & $a_{z2}$ & $a_{z3}$ & $a_{z4}$ & $RSS$ \\
	      &           & $[\mu m^2]$ & $[\mu m^2]$ & $[\mu m^2]$ & $[\mu m^2]$ & $[nm^2]$ \\
	    \hline
		liter. & $4.59423$ & $0.06206$ & $0.04763$ & $110.807$ & $86.122$ & $1910$ \\
		comp.   & $4.59423$ & $0.06272$ & $0.04814$ & $-$ & $-$ & $257$ \\
		uncer.& $0.00015$ & $0.0004$ &  $4.5\times 10^{-6}$ & $-$ & $-$ & $-$\\  \hline
		\end{tabular}
	\caption{Comparison of the literature Ref.~\cite{Kato2002} and calculated Sellemeier coefficients as in \eqref{eq:Sellmeier}}.
	\label{tab:para}
\end{table}

%\section{Outlook}
		
Summarizing, the presented method of obtaining Sellmeier coefficient is very accurate. Our method was demonstrated for a PPKTP crystal, for which we got the value differing by $113$ standard deviations from the initial one.  The method can be used for any other. Moreover the same approach can be used for computation of thermal coefficients i.e.~ones describing change of refractive index with temperature.  We ran such simulations for $n_z$, using model and thermal coefficients from the same article as before Ref.~\cite{Kato2002}. We fitted two out of four parameters to $349$ data points. As a result we achieved fit with slightly smaller $RSS$: $2.39579 \times 10^{3}$ nm$^2$ versus $2.39612 \times 10^{3}$  nm$^2$, which corresponds to $2.6201$ nm and $2.6202$ nm, respectively. Such small difference stems from $16\%$ change of non-dispersive term in equation describing change of $n_z$ with the temperature.  Further analysis is required in order to investigate a phase matching dependence on the remaining system parameters. 

There is also another way to fully utilize our method. In presented work we measured first order quasi phase matching, which led to obtaining very accurate data about Sellmeier coefficients which dominate dispersion in visible range. Measuring higher order quasi phase matching might lead to obtaining additional information about dispersion in crystal, allowing for wider characterization. Our calculation with new set of Sellmeier coefficients implicates that for the same setup we should be able to observe second order quasi phase matching. In our setup, the visible photons with wavelength varying from $428$ nm to $439$ nm for pump wavelength range $390$ nm to $400$ nm should be exiting the crystal at opening angle $\Theta_{VIS}=9$ degree. These wavelengths imply that infrared photon wavelength should be ranging from around $4378$ nm to $4485$ nm. This quasi phase matching would not only depend on all five Sellmeier coefficient describing $n_z$ but also, because of larger angle, would carry same information about other parameters.

%In order to make sure, the observed change in value is connected to change in parameters describing refractive index  rather than the ones describing dependence of periodic polling length further analysis is required.

%\comp{Kolejny rzad QPM -- komentarz}

%\section*{Funding Information}
%The authors acknowledge support by the National Laboratory FAMO in Torun, Poland, financial support by Foundation for Polish Science under Homing Plus no.~2013-7/9 program supported by European Union under PO IG project and by Polish Ministry of Science and Higher Education under grant 6576/IA/SP/2016 and Iuventus Plus grant no.~IP2014 020873. 

{\bf Funding.} Foundation for Polish Science (FNP) (project First Team co-financed by the European Union under the European Regional Development Fund, project Homing Plus grant no.~2013-7/9); Ministry of Science and higher Education, Poland (MNiSW) (grant no.~6576/IA/SP/2016, grant no.~ IP2014 020873); National Science Center, Poland (NCN) (Sonata 12 grant no.~2016/23/D/ST2/02064)

{\bf Acknowledgement.} The authors thank Marcin Bober, Mateusz Borkowski, Roman Ciurylo, Michal Zawada for insightful discussions and National Laboratory of Atomic, Molecular and Optical Physics, Torun, Poland for a support.

{ $^\dagger$ These authors contributed equally to this work.}

\end{document}